\documentclass[12pt]{article}
\usepackage{graphicx}
\begin{document}
\begin{center}
{\bf{$(\alpha ')^4$ CORRECTIONS IN HOLOGRAPHIC LARGE $N_c$ QCD
 AND $\pi - \pi$ SCATTERING}}
\end{center}

\vspace{1.0cm}

\begin{center}
R.Parthasarathy{\footnote{e-mail: sarathy@cmi.ac.in}} \\
Chennai Mathematical Institute \\
Plot Number  H1 SIPCOT IT Park, Padur Post, Siruseri 603 103, India \\
and \\
K.S.Viswanathan{\footnote{e-mail: kviswana@sfu.ca}} \\
Department of Physics, Simon Fraser University and IRMACS Centre \\
Burnaby, BC., V5A 1S6, Canada. \\
\end{center} 

\vspace{1.0cm}

{\noindent{\it{Abstract}}}

\vspace{0.5cm}

We calculate the ${\alpha'}^4$ corrections to the non-Abelian DBI action on the $D8$-brane in the 
holographic dual of large $N_c$ QCD proposed by Sakai and Sugimoto. These give rise to higher 
derivative terms, in particular, four derivative contact terms for the pion field with the coupling uniquely 
determined. We calculate the pion-pion scattering amplitude near threshold. The results respecting   
unitarity are in qualitative agreement with the experimental values. 

\newpage

A holographic dual of QCD with $N_f$ massless quarks using a $D4/D8$-brane configuration in Type-II-A string theory, within the framework of AdS/CFT, has been proposed by Sakai and Sugimoto [1,2]. 
This describes the low energy phenomena of  large $N_c$ 
QCD such as the spontaneous breaking of chiral 
$U(N_f)_L\times U(N_f)_R$ symmetry  
to the diagonal subgroup $U(N_f)_{L+R}$. In this model, the ingredients 
are an $U(N_f)$ five-dimensional Yang-Mills and Chern-Simons theory on a curved background, 
originating from the low energy effective action on the probe $D8$-branes embedded into the $D4$ 
background [3]. The entire tower of vector mesons and the pions are in a single $U(N_f)$ gauge field 
in five dimensions, simplifying possible interaction structures among mesons. 

\vspace{0.5cm}

In the AdS/CFT correspondence [4,5,6] scenario, the baryons are constructed by $D4$-branes wrapped 
on non-trivial cycles [7,8]. In the $D4/D8$ model [1,2], baryons are identified as $D4$-branes 
wrapped on a non-trivial 4-cycle in the $D4$ background, realising $D4$-brane as a small instanton 
configuration in the world-volume gauge theory on the probe $D8$-brane. The pion effective action 
obtained from the 5-dimensional YM theory is the Skyrme action [9] in which baryons appear as 
solitons, with the identification of the baryon number as a winding number  and 
equivalent to the instanton number in the 5-d YM theory. 

\vspace{0.5cm}

The chiral lagrangian derived from the $D4/D8$-brane model [1,2] is found to describe the axial coupling $g_A$ and the electromagnetic form factors of the nucleon [10,11]. Using the hedgehog 
ansatz [9,12] for the chiral field in the Skyrme model, the mass and the root-mean-square radius of the 
brane-induced Skyrmion compare well with the standard Skyrmion [13].  In the studies of the $D4/D8$ 
model [1,2] thus far, only the leading terms in the non-Abelian Dirac-Born-Infeld (DBI) action of the 
$D8$-brane, i.e., order of $(2\pi\alpha ')^2$, have been   considered to obtain a 4-dimensional action 
for pions, although higher order corrections have been considered in the fluctuation analysis in 
background intersecting branes [14] and in the computation of soliton mass [11]. 

\vspace{0.5cm}

In this work, we include higher-order terms of the DBI-action up to $(2\pi \alpha')^4$ to obtain a 
5-dimensional $D8$-brane action. 
 When this is applied to pions, we obtain, besides the Skyrme 
action, two terms involving four pion fields, {\it{precisely}} of the form suggested by Weinberg [15] in 
his phenomenological approach, thus bringing the holographic dual QCD in closer connection with the 
realistic QCD. The coupling strength of these terms are dimensionless and their values uniquely 
determined.  
 It is to be noted that while in the phenomenological lagrangian approach, it was 
necessary to arrange all the pion couplings as derivative interactions, to suppress the incalculable 
graphs in which soft pions would be emitted from internal lines of a hard particle processes, in our case,
all these pion couplings naturally come as derivative interactions.  As an application, we have evaluated the $\pi-\pi$ scattering amplitude $R^0_0$ defined in Sannino and Schechter [16] and find that the inclusion of the higher order derivative 
terms ameliorate the $R^0_0$ curve, avoiding violation of unitarity, bringing it closer to  
Roy's curve [17] as given in [16].  Within the ${\alpha'}^2$ terms of the $D8$-brane DBI action, it was 
shown in [2] that when an infinite tower of massive vector mesons are included, the low energy 
$\pi-\pi$ scattering is governed only by the chiral lagrangian  for pions. We are here including the 
${\alpha'}^4$ corrections from the $D8$-brane DBI action to the chiral lagrangian for pions. 

 \vspace{0.5cm}

 Briefly, in the SS model [1], the $D4$ background consisting of $N_c$ flat $D4$-branes with one of the 
 spatial world-volume directions $\tau$ compactified on $S^1$, is given by the supergravity solution 
 [18]
 \begin{eqnarray}
 (ds)^2={\Big(\frac{U}{R}\Big)}^{\frac{3}{2}}({\eta}_{\mu\nu}dx^{\mu}dx^{\nu}+f(U)d{\tau}^2)+
 {\Big(\frac{R}{U}\Big)}^{\frac{3}{2}}(\frac{dU^2}{f(U)}+U^2d{\Omega}^2_4), \nonumber \\
 e^{\phi}=g_s{\Big(\frac{U}{R}\Big)}^{\frac{3}{4}};F_4=\frac{2\pi N_c}{V_4}{\epsilon}_4; f(U)=
 1-\frac{U_{KK}^3}{U^3},
 \end{eqnarray}
 where $x^{\mu}(\mu=0,1,2,3) $and $\tau$ are the directions along which the $D4$-brane is extended,
 $d{\Omega}^2_4, {\epsilon}_4, V_4=\frac{8{\pi}^2}{3}$ are the line element, volume form and the 
 volume of unit $S^4$, $R$ and $U_{KK}$ are parameters, the coordinate $U$ is bounded from 
 below ($U\geq U_{KK}$), $U=U_{KK}$ corresponds to a horizon in the supergravity solution, 
 $g_s(=e^{<\phi>})$ is the string coupling constant and $R^3=\pi g_s N_c {\ell}^3_s$, ${\ell}_s$ being the 
 string length.  The 4-form  is $F_4=dC_3=\frac{2\pi N_c}{V_4}{\epsilon}_4$ with ${\epsilon}_4$ as the 
 volume form of $S^4$.
 With $\tau$ periodic, the conical singularity at $U=U_{KK}$ is avoided by having the 
 period $\delta \tau$ of $\tau$ as $\frac{4\pi}{3}\frac{R^{\frac{3}{2}}}{U^{\frac{1}{2}}_{KK}}$. The 
 Kaluza-Klein mass scale is $M_{KK}=\frac{2\pi}{\delta \tau}=\frac{3}{2}\frac{U^{\frac{1}{2}}_{KK}}{R
 ^{\frac{3}{2}}}$, below which the dual gauge theory is effectively the same as 4-dimensional YM 
 theory, with $g_{YM}^2=4{\pi}^2g_sl_s/{\delta \tau}$. Then the parameters $R,U_{KK}$ and $g_s$ are
 \begin{eqnarray}
 R^3=\frac{1}{2}\frac{g_{YM}^2N_c{\ell}_s^2}{M_{KK}}&;&U_{KK}=\frac{2}{9}g_{YM}^2N_cM_{KK}
 {\ell}_s^2; \nonumber \\
 g_s=\frac{g_{YM}^2}{2\pi M_{KK}{\ell}_s}&;&M_{KK}=\frac{3U^{\frac{1}{2}}_{KK}}{2
 R^{\frac{3}{2}}}.
 \end{eqnarray}
 
\vspace{0.5cm}

The induced metric on the $D8$-brane, embedded in the $D4$-background (1) with $U=U(\tau)$ is 
\begin{eqnarray}
(ds)^2_{D8}={\Big(\frac{U}{R}\Big)}^{\frac{3}{2}}{\eta}_{\mu\nu}dx^{\mu}dx^{\nu}+\{
{\Big(\frac{U}{R}\Big)}^{\frac{3}{2}}f(U)+{\Big(\frac{R}{U}\Big)}^{\frac{3}{2}}\frac{{U'}^2}{f(U)}\}d{\tau}^2 +
{\Big(\frac{R}{U}\Big)}^{\frac{3}{2}}U^2d{\Omega}^2_4,
\end{eqnarray}
where $U'\equiv \frac{dU}{d\tau}$. The $N_f$ $D8-{\bar{D8}}$ pairs are separately placed along the 
anti-podal points of $\tau$ (see figures 1,2 of [13]) and are smoothly interpolated with each other. Then,
the $U(N_f)_{D8}\times U(N_f)_{\bar{D8}}$ gauge symmetry breaks in to $U(N_f)$ gauge symmetry 
which is interpreted as a holographic manifestation of chiral symmetry breaking in QCD, realizing 
the chiral symmetry breaking by the geometrical connection of $D8$ and $\bar{D8}$ branes. As the 
probe $D8$-brane world volume is on a plane of constant $\tau$, the induced metric (3) can be 
written as
\begin{eqnarray}
(ds)^2_{D8}={\Big(\frac{U(z)}{R}\Big)}^{\frac{3}{2}}{\eta}_{\mu\nu}dx^{\mu}dx^{\nu}+\frac{4}{9}
{\Big(\frac{R}{U(z)}\Big)}^{\frac{3}{2}}\frac{U_{KK}}{U(z)} dz^2+{\Big(\frac{R}{U(z)}\Big)}^{\frac{3}{2}}
U^2(z) d{\Omega}^2_4,
\end{eqnarray}
with $U^3=U^3(z)=U^3_{KK}+U_{KK}z^2$. The $D8$-brane extends along $x^{\mu}(\mu=0,1,2,3)$ 
and $z$ directions, wrapping around $S^4$.

\vspace{0.5cm}

The gauge field on the probe $D8$-brane has nine components, $A_{\mu}, A_z$ and $A_i$ (i=
5,6,7,8, the coordinates on $S^4$).   The non-Abelian DBI-action on 
$D8$-brane is 
\begin{eqnarray}
S^{DBI}_{D8}=T_8\int d^9x e^{-\phi}\  STr \sqrt{-det(g_{MN}+(2\pi\alpha ')F_{MN})},
\end{eqnarray}
where $\alpha'={\ell}_s^2$ is the Regge slope parameter and $T_8=\frac{1}{(2\pi)^8{\ell}_s^9}$ is the 
surface tension of the $D8$-brane. $F_{MN}={\partial}_MA_N-{\partial}_NA_M +i[A_M,A_N]$, 
$g_{MN}$ is the induced metric on $D8$-brane given in (4) and $M,N$ take values $(0,1,2,3,\cdots 8)$.
 In (5) $STr$ is the symmetric trace. 
From (4), we have $e^{-\phi}
\sqrt{-det g_{MN}}=\frac{2}{3}R^{\frac{3}{2}}U_{KK}^{\frac{1}{2}}U^2(z)g_s^{-1}$. The gravitational 
energy of the $D8$-brane in general coordinates is $S^{DBI}_{D8}|_{A_M=0}$ and subtracting the 
gravitational energy as a vacuum energy relative to the gauge sectors, 
\begin{eqnarray}
S^{DBI}_{D8}-S^{DBI}_{D8}|_{A_M=0}=T_8\int d^9x e^{-\phi}\ STr
 \{ \sqrt{-det(g_{MN}+(2\pi \alpha')F_{MN})}
-\sqrt{-det g_{MN}}\}. 
\end{eqnarray}
This is expanded as in [19] to give (we denote the left side of (6) by ${\tilde{S}}^{DBI}_{D8}$)
\begin{eqnarray}
{\tilde{S}}^{DBI}_{D8}=\frac{T_8}{4}(2\pi \alpha')^2\int d^9x e^{-\phi}\sqrt{-det g_{MN}}\  Tr[ F_{MN}
F^{MN} \nonumber \\
-\frac{1}{3}(2\pi \alpha')^2\{ F_{MN}F^{RN}F^{ML}F_{RL}+\frac{1}{2}F_{MN}F^{RN}F_{RL}F^{ML}
\nonumber \\
-\frac{1}{4} F_{MN}F^{MN}F_{RL}F^{RL}-\frac{1}{8}F_{MN}F^{RL}F^{MN}F_{RL}\} 
+O({\alpha '}^4)]. 
\end{eqnarray}
Now restricting to $SO(5)$ singlets, we set $A_i=0$ and take $A_{\mu}$ and $A_z$ to be 
independent of the coordinates of $S^4$. The integration over the $S^4$ coordinates is performed.
Then,  the full $D8$-brane DBI action up to 
$(\alpha')^4$ terms becomes,
\begin{eqnarray}
{\tilde{S}}^{DBI}_{D8}&=&\Big( \frac{2}{3} T_8 R^{\frac{3}{2}}U_{KK}^{\frac{1}{2}}V_4 g_s^{-1}\Big) 
(2\pi \alpha ')^2 \int d^4x\  dz\  2Tr\  [\frac{1}{4}\frac{R^3}{U(z)}{\eta}^{\mu\nu}{\eta}^{\lambda\sigma}
F_{\mu\lambda}F_{\nu\sigma} \nonumber \\
&+&\frac{9}{8}\frac{U^3(z)}{U_{KK}}{\eta}^{\mu\nu}F_{\mu z}F_{\nu z} 
\nonumber \\
&-&\frac{1}{12}(2\pi \alpha')^2\{ \frac{R^6}{U^4(z)}{\eta}^{\mu\nu}{\eta}^{\lambda\sigma}
{\eta}^{\rho\delta}{\eta}^{\alpha\beta}\Big( F_{\mu\lambda}F_{\delta\sigma}F_{\nu\beta}
F_{\rho\alpha} -\frac{1}{8}F_{\mu\lambda}F_{\beta\delta}F_{\nu\sigma}F_{\alpha\rho}   \nonumber \\
&+&\frac{1}{2}F_{\mu\lambda}F_{\beta\sigma}F_{\alpha\rho}F_{\nu\delta}-\frac{1}{4}F_{\mu\lambda}
F_{\nu\sigma}F_{\alpha\rho}F_{\beta\delta}\Big) \nonumber \\
&+&\frac{9}{4}\ \frac{R^3}{U_{KK}}\ {\eta}^{\mu\nu}{\eta}^{\lambda\sigma}{\eta}^{\alpha\beta}\Big(
2F_{\mu\lambda}F_{\beta\sigma}F_{\nu z}F_{\alpha z}+F_{\mu\lambda}F_{\sigma z}F_{\nu\beta}
F_{\alpha z} 
+F_{\mu z}F_{\sigma \nu}F_{\beta z}F_{\lambda\alpha} \nonumber \\
&-&\frac{1}{2}F_{\mu\lambda}F_{\beta z}
F_{\nu\sigma}F_{\alpha z}+\frac{1}{2}F_{\mu\lambda}F_{\beta\sigma}F_{\alpha z}F_{\nu z} 
+\frac{1}{2}F_{\mu\lambda}F_{\sigma z}F_{\alpha z}F_{\nu \beta} \nonumber \\
&+&\frac{1}{2}F_{\mu z}F_{\beta z}F_{\alpha\lambda}F_{\nu\sigma}+\frac{1}{2}F_{\beta z}
F_{\mu z}F_{\nu \sigma}F_{\alpha\lambda}-F_{\mu\lambda}F_{\nu\sigma}F_{\alpha z}
F_{\beta z}\Big) \nonumber \\
&+&\frac{81}{16}\ \frac{U^4(z)}{U_{KK}^2}{\eta}^{\mu\nu}{\eta}^{\lambda\sigma}\Big( \frac{1}{2}
F_{\mu z}F_{\sigma z}F_{\nu z}F_{\lambda z}+F_{\mu z}F_{\sigma z}F_{\lambda z}F_{\nu z}
   \Big)\}],
\end{eqnarray}
where we follow the normalization $Tr (T^aT^b)=\frac{1}{2}
{\delta}^{ab}$. 

\vspace{0.5cm}

The action in (8) is general up to ${\alpha '}^4$ terms. The 5-dimensional gauge fields  
$A_{\mu}(x,z)$ and $A_z(x,z)$ can 
be  expanded using complete sets of functions of $z$ and a 4-dimensional action can be obtained by 
integrating  over $z$. We use the gauge  $A_z=0$  
(see [1] and [13] for details of realization of this gauge choice). We are interested in the pions only 
and so we expand as in [1],
\begin{eqnarray}
A_{\mu}(x,z) &=& U^{-1}(x){\partial}_{\mu} U(x)\ {\psi}_+(z),
\end{eqnarray}
where ${\psi}_+(z)=\frac{1}{2}+\frac{1}{\pi}tan^{-1}(\frac{z}{U_{KK}})$ and $U(x)=e^{\frac{2i}{f_{\pi}}\pi(x^
{\mu})}$, with $f_{\pi}$ as a parameter (at this stage) and $\pi(x^{\mu})$ is the pion field.   The function 
${\psi}_+(z)$ is closely related to the implementation of the gauge $A_z=0$ [1]. From (9) it follows that 
\begin{eqnarray}
F_{\mu\nu}&=&[U^{-1}{\partial}_{\mu}U,U^{-1}{\partial}_{\nu}U]\ {\psi}_+(z)({\psi}_+(z)-1), \nonumber \\
F_{z\mu}&=&U^{-1}{\partial}_{\mu}U\ ({\partial}_z{\psi}_+(z))\ \equiv \ U^{-1}{\partial}_{\mu}U 
{\hat{\phi}}_0(z),
\end{eqnarray}
where ${\hat{\phi}}_0(z)\ =\ \frac{U_{KK}^2}{\pi}\ \frac{1}{U^3(z)}$ where $U(z)$ is defined below (4). 
  Substituting (10) in (8), we encounter 
the following $z$-integrals which are numerically evaluated. 
\begin{eqnarray}
\frac{R^3}{4}\int_{-\infty}^{\infty} \frac{1}{U(z)} {\psi}_+^2({\psi}_+-1)^2\ dz &=& \frac{R^3}{4{\pi}^4}\times 
15.2463,
\end{eqnarray}
\begin{eqnarray}
\frac{9}{8U_{KK}}\int_{-\infty}^{\infty}U^3(z){\hat{\phi}}_0^2(z)dz&=& \frac{9U_{KK}}{8\pi}, 
\end{eqnarray}
\begin{eqnarray}
\int_{-\infty}^{\infty}U^4(z){\hat{\phi}}_0^4(z)dz&=& \frac{U_{KK}}{{\pi}^4}\times 1.275,
\end{eqnarray}
\begin{eqnarray}
\int_{-\infty}^{\infty}{\psi}_+^2({\psi}_+-1)^2{\hat{\phi}}_0^2(z)dz &=& \frac{1}{U_{KK}{\pi}^6}\times 7.4545,
\end{eqnarray}
\begin{eqnarray}
\int_{-\infty}^{\infty}\frac{1}{U^4(z)}{\psi}_+^4({\psi}_+-1)^4 dz&=& \frac{1}{U_{KK}^3{\pi}^8}\times 43.7376.
\end{eqnarray}

\vspace{0.5cm}

Then, the $D8$-brane DBI action in four dimensions becomes
\begin{eqnarray}
S&=&\int d^4x Tr \{\frac{f_{\pi}^2}{4} L_{\mu}L^{\mu}+\frac{1}{32e^2}[L_{\mu},L_{\nu}]^2\} \nonumber \\
&+&{\tilde{T}}_8(2\pi \alpha')^2\Big( -\frac{1}{12}(2\pi \alpha')^2\Big)\int d^4x Tr \  [
 \Big(\frac{2R^6}{{\pi}^8U_{KK}^3}\times 43.7376\Big) \nonumber \\
& &{\eta}^{\mu\nu}{\eta}^{\lambda\sigma}
{\eta}^{\rho \delta}{\eta}^{\alpha\beta} \{ [L_{\mu},L_{\lambda}] [L_{\delta},L_{\sigma}] [L_{\nu},
L_{\beta}] [L_{\rho},L_{\alpha}]  \nonumber \\
&-&\frac{1}{8}[L_{\mu},L_{\lambda}] [L_{\beta},L_{\delta}] [L_{\nu},L_{\sigma}] [L_{\alpha},L_{\rho}] 
\nonumber \\
&+& 
\frac{1}{2}[L_{\mu},L_{\lambda}] [L_{\beta},L_{\sigma}] [L_{\alpha},L_{\rho}] [L_{\nu},L_{\delta}] \nonumber 
\\
&-&\frac{1}{4}[L_{\mu},L_{\lambda}] [L_{\nu},L_{\sigma}] [L_{\alpha},L_{\rho}] [L_{\beta},L_{\delta}] \}
\nonumber \\
&+&\Big(\frac{9R^3}{2U_{KK}^2{\pi}^6}\times 7.4545\Big){\eta}^{\mu\nu}{\eta}^{\lambda\sigma}
{\eta}^{\alpha\beta}\{2[L_{\mu},L_{\lambda}] [L_{\beta},L_{\sigma}] L_{\nu}L_{\alpha} \nonumber \\
&+&[L_{\mu},L_{\lambda}] L_{\sigma} [L_{\nu},L_{\beta}] L_{\alpha}+L_{\mu} [L_{\sigma},L_{\nu}]
L_{\beta} [L_{\lambda},L_{\alpha}] \nonumber \\
&-&\frac{1}{2}[L_{\mu},L_{\lambda}] L_{\beta} [L_{\nu},L_{\sigma}] L_{\alpha}+\frac{1}{2}
[L_{\mu},L_{\lambda}] [L_{\beta},L_{\sigma}] L_{\alpha}L_{\nu} \nonumber \\
&+&\frac{1}{2}[L_{\mu},L_{\lambda}] L_{\sigma}L_{\alpha} [L_{\nu},L_{\beta}]+\frac{1}{2}L_{\mu}
L_{\beta}[L_{\alpha},L_{\lambda}] [L_{\nu},L_{\sigma}] \nonumber \\
&+& \frac{1}{2} L_{\beta} L_{\mu} [ L_{\nu},L_{\sigma}] [L_{\alpha},L_{\lambda}] - [L_{\mu},
L_{\lambda}] [L_{\nu},L_{\sigma}] L_{\alpha} L_{\beta} \} \nonumber \\
&+&\Big( \frac{81\times 1.275}{8U_{KK}{\pi}^4}\Big)
{\eta}^{\mu\nu}{\eta}^{\lambda\sigma}\{ \frac{1}{2}
L_{\mu}L_{\sigma}L_{\nu}L_{\lambda}+L_{\mu}L_{\nu}L_{\sigma}L_{\lambda}
 \} ],
\end{eqnarray}
where ${\tilde{T}}_8=\frac{2}{3}R^{\frac{3}{2}}\ U_{KK}^{\frac{1}{2}}\ T_8\ V_4\ g_s^{-1}$ , $L_{\mu}
\ =\ U^{-1}{\partial}_{\mu}U$ 
and 
\begin{eqnarray}
{\tilde{T}}_8(2\pi \alpha')^2&=& \frac{\pi f_{\pi}^2}{9U_{KK}} , \nonumber \\
                                                &=& \frac{1}{32e^2}\ \frac{2{\pi}^4}{15.2463\times R^3}.                                                
\end{eqnarray}

\vspace{0.5cm}

For $N_f=2$,  a lagrangian describing massless pions up to ${\alpha '}^4$ and four derivatives of 
$U(x)$ is given by the first two and the last two terms in (16) which should describe the properties of pion
and $\pi-\pi$ scattering. The first two terms reproduce the Skyrme model and the static properties of the 
pion are well described by the hedgehog ansatz [12]. The second term in (16) is the familiar Skyrme 
term introduced by Skyrme to stabilize the soliton. The holographic dual model [1] has this term 
naturally. When massive vector mesons (infinite tower) are introduced, as said in the beginning, the 
contribution from the Skyrme term for $\pi-\pi$ scattering gets cancelled and the resulting lagrangian for 
the pions is just the chiral lagrangian [2]. Thus the pion-pion scattering here will be described by the first 
and the last two terms in (16) which are precisely the terms in the phenomenological lagrangian of 
Weinberg [15], with the coefficients (coupling constants) fixed by the parameters of the holographic 
model.  The sixth and eighth derivative terms in (16) will contribute to $\pi\pi$ scattering leading to four 
and six pions. 

\vspace{0.5cm}

Now we consider the $\pi \pi\rightarrow \pi\pi$ scattering.  Weinberg [20] obtained the $\pi-\pi$ scattering amplitude  for 
${\cal{L}}=-\frac{f_{\pi}^2}{4} Tr(L_{\mu}L^{\mu})$ (the first term in (16)) as 
\begin{eqnarray}
A(s,t,u)&=& \frac{s}{f_{\pi}^2},
\end{eqnarray}
where $s,t,u$ are the Mandelstam variables, $s+t+u=0$.  In  holographic QCD, pion mass 
can be realized by introducing instantons on the $S^4$ [21] which will not affect (16) except for a mass 
term for the pions.  Then, (18) reads as $A(s,t,u)=\frac{s-m_{\pi}^2}{f_{\pi}^2}$ with $s+t+u=4m^2_{\pi}$. 
Sannino and Schechter 
[16] found that the dependence of $R_0^0(s)$ on $\sqrt{s}$ 
did not follow the Roy curves [17] 
and violated unitarity. Following the phenomenological lagrangian of Weinberg [15], they [16] 
introduced four-derivative contact terms (which are the last two terms in (16)) with arbitrary coefficients, 
and adjusted them so as to have vanishing contribution from these for threshold scattering.  Notice that 
 in the 
holographic model,  the couplings are fixed uniquely and they involve only the Yang-Mills coupling 
$^2_{YM}$. 
Further details of 
pion-pion scattering can be found in [21,22].  The chiral lagrangian for pions in the holographic model, 
up to ${\alpha'}^4$ corrections and up to four derivatives are given from (16) as 
\begin{eqnarray}
{\cal{L}}_{eff}^{pion}&=& \frac{f_{\pi}^2}{4} Tr ({\partial}_{\mu}U\ {\partial}^{\mu}U^{\dagger}) \nonumber \\
& -& C_4\ Tr \{ \frac{1}{2}{\partial}_{\mu}U\ {\partial}_{\nu}U^{\dagger}\ {\partial}^{\mu}U\  {\partial}^{\nu}U
^{\dagger}\ +\ {\partial}_{\mu}U\ {\partial}^{\mu}U^{\dagger}\ {\partial}_{\nu}U \ {\partial}^{\nu}
U^{\dagger}\}.
\end{eqnarray}
The $(\alpha ')^4$ corrections give the four derivative contact interaction in (19) with the dimensionless 
 coupling constant $C_4$  as 
\begin{eqnarray}
C_4&=& 
{\tilde{T}}_8(2\pi \alpha')^4\ \frac{81\times 1.275}{96\ U_{KK}\ {\pi}^4}
=\frac{1.173\times 10^{-3}}{g_{YM}^2} ,
\end{eqnarray}
using (2) and (17). 

\vspace{0.5cm}

The pion-pion scattering amplitude [16] from (19) is 
\begin{eqnarray}
A(s,t,u)&=& \frac{s-m_{\pi}^2}{f_{\pi}^2}-\frac{2C_4}{f_{\pi}^4}\{ (t-2m_{\pi}^2)^2+(u-2m_{\pi}^2)^2 
\nonumber \\
&+&(s-2m_{\pi}^2)^2\},
\end{eqnarray}
for which the partial wave amplitude $T^0_0(s)$ is 
\begin{eqnarray}
T^0_0(s)&=&\frac{1}{64\pi}\sqrt{1-\frac{4m_{\pi}^2}{s}}\times \int_{-1}^1 d\ cos{\theta}\ T^0(s,t,u),
\end{eqnarray}
where $T^0(s,t,u)\ =\ 3A(s,t,u)+A(t,s,u)+A(u,t,s)$ with
 $s=4({\vec{p}}\ ^2+m_{\pi}^2); \ t=-2{\vec{p}}\ ^2
(1-cos{\theta}); \ u=-2{\vec{p}}\ ^2(1+cos{\theta})$,
 ${\vec{p}}$ the 3-momentum and $\theta$ the 
scattering angle of the pion. Then it is straightforward to obtain 
\begin{eqnarray}
R^0_0(s)=T^0_0(s)&=& \frac{1}{64\pi}\sqrt{1-\frac{4m_{\pi}^2}{s}}\ [ \frac{2}{f_{\pi}^2}(2s-m_{\pi}^2)
\nonumber \\
&-&\frac{10 C_4}{f_{\pi}^4}\{ 2(s-2m_{\pi}^2)^2+s^2+\frac{1}{3}(s-4m_{\pi}^2)^2\} ].
\end{eqnarray}

\vspace{0.5cm}

In Figure.1, we have plotted $R_0^0(s)$ as a function of $\sqrt{s}$ with 
and without ${\alpha'}^4$ corrections using $f_{\pi}=95 MeV$ and for two representative values of 
$g_{YM}^2=4\pi {\alpha}_s$. Curve I is without the ${\alpha'}^4$ corrections. Curve II is with the 
${\alpha'}^4$ corrections using $g^2_{YM}=4\pi {\alpha}_s$ with  
the value of ${\alpha}_s=0.12$ at the Z-boson mass [23].  This value for $g^2_{YM}$ is for real QCD at 
short distances. In view of the holographic model used here, it will be consistent to use large $N_c$ 
value for $g^2_{YM}$. By writing $g^2_{YM}$ as $\frac{(\lambda N_c)}{N_c}$, where $\lambda = 
g^2_{YM}N_c$, the 'tHooft coupling parameter, we adopt the fit for  $g_A$ in [10] with $\lambda N_c 
\simeq 26$ and shift the denominator $N_c$ by $N_c+2$ following Dashen and Manohar [24]. The 
amplitude $R^0_0$ with these is displayed in Figure.1 as curve III. 
From the figure, it is seen that the ${\alpha'}^4$ 
corrections are important to be consistent with the unitarity.  Curve III further respects the unitarity 
bound $|R^0_0|\leq \frac{1}{2}$. The numerical values for $R(s)$ are in qualitative agreement with the 
real part of the $I=0;\ell =0$ partial wave amplitude using the phase shifts given in [22]. We find the 
$I=1$ amplitude $T^1(s,t,u)=A(t,s,u)-A(u,t,s)$, using (21) is independent of the ${\alpha'}^4$ 
corrections. The $I=2$ amplitude $A^2(s,t,u)=A(t,s,u)+A(u,t,s)$ is used to calculate the partial wave 
amplitudes $R^2_0(s)$ and $R^2_2(s)$ for $\ell=0,2$ respectively. It is noted that the $I=2; \ell=2$ 
partial wave scattering amplitude $R^2_2(s)$ involves only the ${\alpha'}^4$ terms after the angular 
integration. The experimental phase shifts from [25] are used to find these amplitudes using $R^2_{\ell}(s)=\frac{1}{2}sin(2{\delta}^2_{\ell})$ with ${\eta}^{(2)}_{\ell}$  the $I=2$ inelasticity parameter set 
equal to unity 
and compared with our theoretical 
values in Tables 1 and 2. As the experimental phase shifts are available over a bin for $\sqrt{s}$, we 
have taken the median values. 

\vspace{0.5cm}

\begin{center}
{\bf{Table. 1}} \\

\vspace{0.3cm}

{\it{The results for $R^2_0(s)$. The second column is our theoretical values and the third colummn is using the phase shifts from [22,25]}}

\vspace{0.3cm}

\begin{tabular}{|l|l|l|}\hline 
$\sqrt{s} $(GeV) & Theory    & $\frac{1}{2}sin(2{\delta}^2_0)$ \\ \hline 
                           &                  &      \\
0.35 & -0.07 &-0.069$\pm$0.035 \\
0.45 &-0.155&-0.137$\pm$0.015 \\
0.60&-0.268 &-0.18$\pm$0.027 \\
0.64&-0.323&-0.26$\pm$0.023 \\ \hline 
\end{tabular}
\end{center}

\vspace{0.5cm}

\begin{center}                           

\vspace{0.3cm}

{\bf{Table.2}}

\vspace{0.3cm} 

{\it{The results for $R^2_2(s)$. The second column is our theoretical values and the third column is 
using the phase shifts from [22,25]}}

\vspace{0.3cm}

\begin{tabular}{|l|l|l|}\hline 
$\sqrt{s} $ (GeV) & Theory &$\frac{1}{2}sin(2{\delta}^2_2)$ \\ \hline 
                            &              &   \\
 0.75 &-0.0082 &-0.015$\pm$0.005 \\
 0.80&-0.011&-0.04$\pm$0.005 \\
 1.0&-0.03 &-0.035$\pm$0.01 \\ \hline 
 \end{tabular}
 \end{center}
 
 \vspace{0.5cm}
 
 The numerical values are in reasonable agreement with the results using the experimental phase shifts.

\vspace{0.5cm}
\newpage
{\noindent{\bf{Acknowledgements}}}

\vspace{0.5cm}

One of us (R.P) acknowledges with thanks the hospitality at the Physics Department and IRMACS,
 Simon Fraser 
University, and the Chennai Mathematical Institute for granting leave. This research is supported by an
operating grant from Natural Sciences and Engineering Council of Canada.  Useful correspondence 
with J.R.Pel\`{a}ez is acknowledged with thanks. 

\vspace{1.5cm}

{\noindent{\bf{References}}}

\vspace{0.5cm}

\begin{enumerate}
\item T.Sakai and S.Sugimoto, Prog.Theor.Phys. {\bf{113}} , 843 (2005); [arXiv:hep-th/0412141].
\item T.Sakai and S.Sugimoto, Prog.Theor.Phys. {\bf{114}}, 1083 (2006); [arXiv:hep-th/0507073].
\item E.Witten, Adv.Theor.Math.Phys. {\bf{2}}, 505 (1998); [arXiv:hep-th/9803131].
\item J.M.Maldacena, Adv.Theor.Math.Phys. {\bf{2}}, 231 (1998); [arXiv:hep-th/9711200].
\item S.S.Gubser,I.R.Klebanov and A.M.Polyakov, Phys.Lett. {\bf{B428}}, 105 (1998); [arXiv:hep-th/9802109]; \\
          E.Witten, Adv.Theor.Math.Phys. {\bf{2}}, 253 (1998); [arXiv:hep-th/9802150].
\item O.Aharony, S.S.Gubser, J.M.Maldacena, H.Ooguri and Y.Oz, Phys.Rep. {\bf{323}}, 183 (2000);
[arXiv:hep-th/9905111].
\item E.Witten, JHEP {\bf{9807}}, 006 (1998); [arXiv:hep-th/9805112].
\item D.J.Gross and H.Ooguri, Phys.Rev. {\bf{D58}}, 106002 (1998); [arXiv:hep-th/9805129]; \\
          C.G.Callan, A.Guijosa and K.G.Savvidy, Nucl.Phys. {\bf{B547}}, 127 (1999); [arXiv:hep-th/9810092].
\item T.H.R.Skyrme, Proc.Roy.Soc.Lond. {\bf{A260}}, 127 (1961); Nucl.Phys. {\bf{31}}, 556 (1962).
\item D.K.Hong,M.Rho,H-U.Yee and P.Yi, "Chiral Dynamics of Baryons from String Theory", arXiv:
hep-th/0701276; "Dynamics of Baryons from String Theory and Vector Dominance", arXiv:
hep-th/07052632. 
\item H.Hata, T.Sakai, S.Sugimoto and S.Yamato,  "Baryons from Instantons in holographic QCD", 
arXiv:hep-th/0701280.
\item A.P.Balachandran,V.P.Nair.S.G.Rajeev and A.Stern, Phys.Rev.Lett. {\bf{49}}, 1124 (1982); \\
          E.Witten, Nucl.Phys. {\bf{B223}}, 422,433 (1983); \\
          G.Adkins, C.Nappi and E.Witten, Nucl.Phys. {\bf{B228}}, 552 (1983). 
\item K.Nawa,H.Suganuma and T.Kojo, "Baryons in holographic QCD", arXiv:hep-th/0612187.
\item S.Nagaoka, Prog.Theor.Phys. {\bf{110}}, 1219 (2004); arXiv: hep-th/0307232. 
\item S.Weinberg, Physica. {\bf{A96}}, 327 (1979).
\item F.Sannino and J.Schechter, Phys.Rev. {\bf{D52}}, 96 (1995).
\item S.M.Roy, Phys.Lett. {\bf{36}}, 353 (1971)
\item M.Kruczenski, D.Mateos, R.C.Myers and D.J.Winters, JHEP {\bf{5}} , 041 (2004); arXiv:
hep-th/0311270.
\item A.A.Tseytlin, Nucl.Phys. {\bf{B501}}, 41 (1997); arXiv:hep-th/9701125.
\item S.Weinberg, Phys.Rev.Lett. {\bf{17}}, 616 (1966).
\item B.Ananthanarayan and P.B\"{u}ttiker,  "Pion-Pion scattering in Chiral Perturbation and Dispersion Relation Theories", arXiv:hep-ph/9902461;\\ 
J.Gasser and H.Leutwyler, Ann.Phys. (N.Y). {\bf{158}}, 142 (1984);
 Nucl.Phys. {\bf{B250}}, 465 
(1985);
 G.Colangelo, J.Gasser and H.Leutwyler, Nucl.Phys. {\bf{B603}}, 125 (2001).
 \item J.R.Pel\`{a}ez and F.J.Yndur\`{a}in, Phys.Rev. {\bf{D71}}, 074016 (2005).
\item "Review of Particle Physics (Particle Data Group)", J.Phys:G, Nucl. Part. Phys. {\bf{33}}, 1 (2006). 
\item R.F.Dashen and A.V.Manohar, Phys.Lett. {\bf{B315}}, 438 (1993); arXiv:hep-ph/9307242.  
\item W.Hoogland et.al., Nucl.Phys. {\bf{B126}}, 109 (1977); \\
          H.J.Losty et.al., Nucl.Phys. {\bf{B69}}, 185 (1974); \\
          N.B.Durusoy et.al., Phys.Lett. {\bf{B45}}, 517 (1973).                                      
 \end{enumerate}  
 
 \vspace{0.5cm}
 
\begin{figure}[ht]
\begin{center}
\includegraphics[bb=4 130 667 678, clip, scale=0.6, keepaspectratio=true]{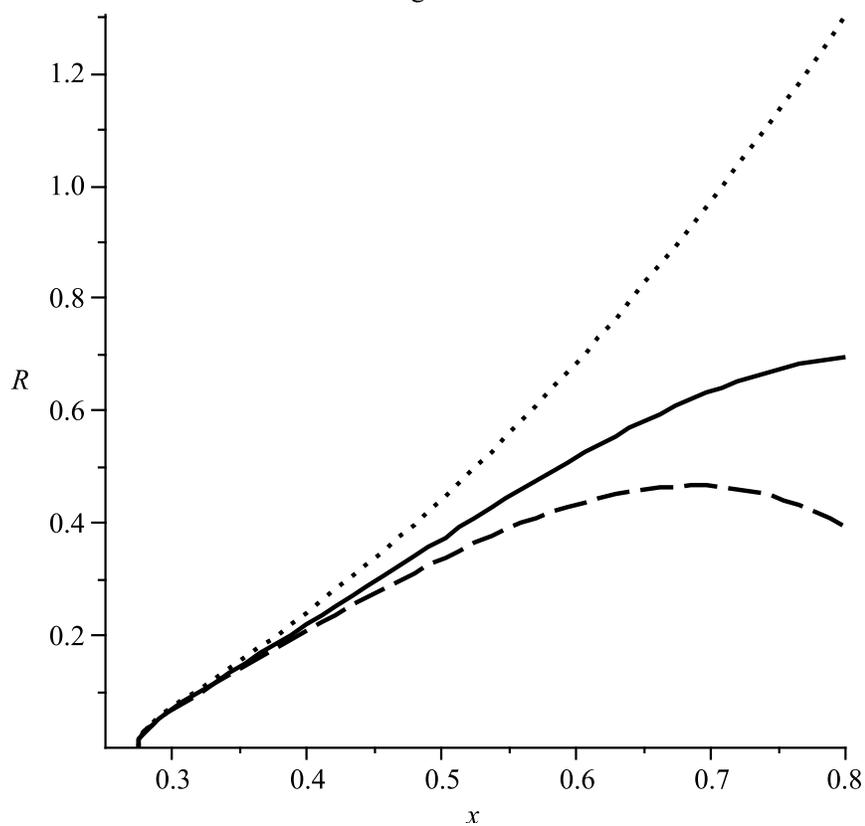}
\caption{$R^0_0(s)=R$ and $x=\sqrt{s}$ in $GeV$.  Curve I is without the ${\alpha '}^4$ corrections. Curves II and III are with these corrections for two representative values of $g^2_{YM}$ (see text).}
\end{center}
\end{figure}

\end{document}